\documentclass[twocolumn, amsmath, amssymb, aps, pra, 10pt]{revtex4-2}
\usepackage[dvipsnames]{xcolor}
\usepackage{graphicx}
\usepackage{enumitem}
\usepackage{siunitx}
\DeclareSIUnit\gauss{G}
\usepackage{booktabs}
\usepackage{multirow}
\usepackage{mathtools}
\usepackage{flushend}
\usepackage[switch]{lineno}
\usepackage[utf8]{inputenc}
  \makeatletter
    \renewcommand\@make@capt@title[2]{%
     \@ifx@empty\float@link{\@firstofone}{\expandafter\href\expandafter{\float@link}}%
      {\textbf{#1}}\@caption@fignum@sep#2\quad}%
    \makeatother
   
\makeatletter 
\renewcommand{\fnum@figure}{\textbf{Fig.~\thefigure}}
\makeatother
\usepackage{braket}

\usepackage{times}
\usepackage{bm}
\usepackage{amsmath}

\newcommand{\probP}{\text{I\kern-0.15em P}}

\newcounter{lastnote}

\begin{document} 



\title{Vector Atom Accelerometry in an Optical Lattice }
\author{Catie LeDesma}
\email[]{Catherine.LeDesma@colorado.edu}

\author{Kendall Mehling}

\author{Murray Holland}

\affiliation{JILA \& Department of Physics, University of Colorado Boulder, CO 80309-0440, USA}

\date{\today}

\begin{abstract}

We experimentally demonstrate two multidimensional atom interferometers capable of measuring both the magnitude and direction of applied inertial forces. These interferometers do not rely on the ubiquitous light-pulses of traditional atom sensors, but are instead built from an innovative design that operates entirely within the Bloch bands of an optical lattice formed by interfering laser beams. Through time-dependent control of the position of the lattice in three-dimensional space, we realize simultaneous Bloch oscillations in two dimensions, and a vector atomic Michelson interferometer. Fits to the observed Bloch oscillations demonstrate the measurement of an applied acceleration of $2g$ along two axes, where $g$ is the average gravitational acceleration at the Earth's surface. For the Michelson interferometer, we perform Bayesian inferencing from a 49-channel output by repeating experiments for selected examples of two-axis accelerations. We demonstrate the resulting accuracy and sensitivity for vector parameter estimation. Our acceleration can be measured from a single experimental run and does not require repeated shots to construct a fringe. We find the performance of our device to be near the quantum limit for the interferometer size and quantum detection efficiency of the atoms. We discuss the reconfigurability of the vector accelerometer and the pathway toward further sensitivity.   

\end{abstract}

\maketitle 


Atom interferometers are exquisite sensors which have been used to perform inertial measurements with ever-increasing precision. These measurements involve forces and accelerations \cite{PhysRevLett.67.181, PhysRevLett.75.2638}, rotations \cite{PhysRevLett.78.2046, 2000CQGra..17.2385G}, and gravity gradients \cite{PhysRevLett.81.971, PhysRevA.65.033608}. They have been used to characterize fundamental constants~\cite{PhysRevLett.70.2706}, to establish certain dark energy constraints \cite{doi:10.1126/science.aaa8883}, and to test the equivalence principle~\cite{PhysRevLett.125.191101}. While atomic interferometers have proven to be excellent precision instruments, the long-standing objective of realizing their sensitivity in field-deployable architectures~\cite{bongs-2019, barrett-2016}  has remained difficult. Scientific endeavors, such as low-orbit measurements of the Earth's gravity field \cite{Tino2019,Lachmann2021} or dark matter detection \cite{Tsai2023}, will benefit from precision atom interferometry~\cite{Tino_2021}. However, these often present dynamically harsh environments and strict requirements on size, weight, and power consumption (SWaP) that are challenging to accommodate for conventional free-fall atom interferometers. 

Our solution to this quantum design problem is Bloch-band interferometry (BBI) in a three-dimensional (3D) optical lattice that uses Bose-Einstein condensed (BEC) atoms as a source~\cite{ledesma2023machinedesigned}. The BBI system offers several advantages over both the traditional pulsed-light atom interferometer, and the shaken-lattice interferometer (SLI)~\cite{PhysRevA.64.063613,PhysRevA.95.043624} where the lattice is shaken in such as way as to allow atoms to propagate through confined states through dynamic tunneling~\cite{PhysRevLett.99.220403}. In BBI, the best aspects of both approaches are combined; the shaking occurs only during the interferometer components, such as beamsplitters and mirrors, thereby allowing the programmability of SLI, but the lattice is fixed in space while the atoms propagate in the conduction band. This means that the atoms transport as essentially free particles rather than tunneling through the lattice allowing large distances to be covered as in a light-pulse interferometer. 

The result is a sensor that is compact with strong (10-100$g$) on-axis restoring forces of the lattice that will permit measurements of inertial signals in environments with inherent thermal and vibrational noise. The programmability aspect allows dynamic configurability of the sensor in software rather than hardware to perform as an accelerometer, gyroscope, or gravity gradiometer in however many dimensions are desired. The atoms can even be frozen in place at their extremum separation while continuously integrating an inertial signal, so that unlike traditional atom interferometers, the sensitivity is not fundamentally limited by the device footprint, i.e., size and shape. Although there are differences, our methodology moves in the same direction as recent hybrid cavity experiments demonstrating coherence times exceeding 1 minute in an optical lattice~\cite{panda2022quantum}, and builds upon early work on component design for atoms in optical lattices~\cite{JHeckerDenschlag_2002}. 

The ability to perform atom interferometry entirely within a three-dimensional optical lattice is enabled by the results of extensive machine-learning and quantum optimal control theory to produce non-intuitive control solutions. We demonstrate that these solutions can be learned offline and implemented in the laboratory with high efficacy. Since our device generates a two-dimensional (2D) diffraction image, we produce a measurement signal with many output channels. This allows us to determine more than one parameter simultaneously from a single image, i.e., a single run of the experiment. We can evaluate the uncertainty of estimated parameters by applying frequency ramps to the lattice lasers, equivalent to applying accelerations to the underlying platform. 

In this paper, we focus on a precision measurement of the vector components of an acceleration~\cite{stolzenberg2024multiaxis, GUO20222291}. We have demonstrated two novel atomic accelerometers capable of measuring both the magnitude and direction of an applied inertial force. Aside from changing the control functions, these interferometers can both be performed without any modifications to the experimental hardware. To our knowledge, we have demonstrated the most sensitive experimental realization of simultaneous Bloch oscillations in 2D, and the first multidimensional Bloch-band Michelson interferometer. Our results highlight the flexibility of a lattice-based architecture for multidimensional sensing \cite{PhysRevLett.111.083001, PhysRevA.109.013327, PhysRevLett.122.043604}, and are a milestone towards realizing metrologically relevant sensitivities in a deployable and reconfigurable atomic-based sensor architecture. 

\vspace*{-1pc}
\subsubsection*{Experimental Procedure}
\vspace*{-.75pc}

\begin{figure}[htb]
    \includegraphics[width=0.3\textwidth]{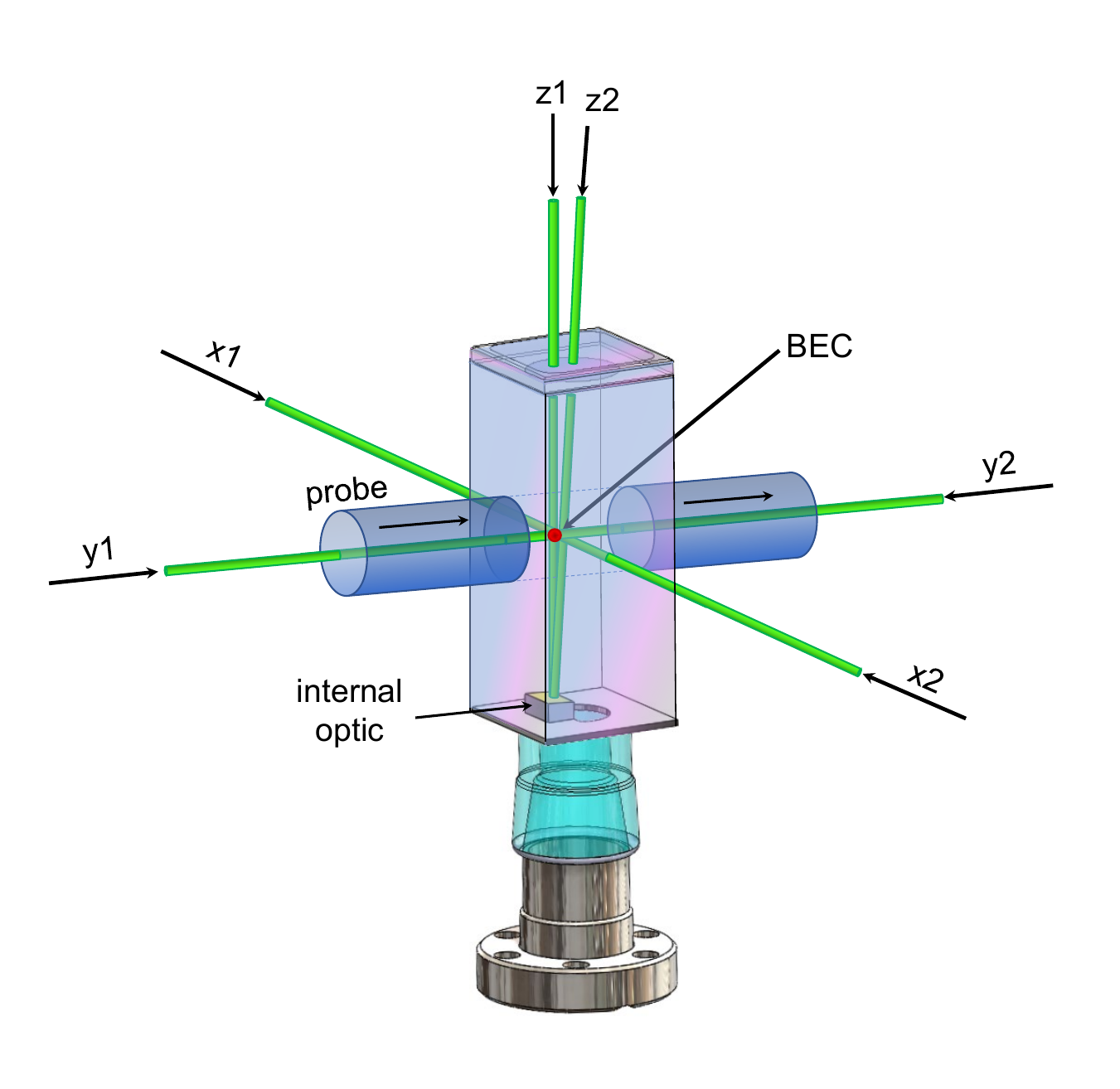}
    \caption{\textbf{Science Chamber:} Vacuum cell showing the geometry of lattice beams (green) oriented along three orthogonal axes. The lattice beams intersect Bose-Einstein condensed atoms (red) over the angled internal optic (gold trapezoid). Although only a single probe beam (blue) is shown, probe beams are aligned to each axis of the lattice to enable imaging from any direction.}
    \label{fig:cell}
\end{figure}

Our all-optical BEC apparatus and procedure has been previously discussed \cite{ledesma2023machinedesigned}, although there have been several additions to incorporate a 3D lattice. As depicted in Fig.~\ref{fig:cell}, a custom anti-reflection (AR) coated internal optic (mirror integrated into the ultrahigh vacuum chamber) is an essential element of the double magneto-optical trap (MOT) system. This allows for a six-beam 3D lattice even though optical access is unavailable on one side. The internal optic is angle polished at 4$^\circ$, allowing projection of two counter-propagating lattice beams from a lattice layer mounted above the experiment. To assist loading of the 3D optical lattice, we optimize the MOT and sub-Doppler cooling stages such that the atoms are located over the internal optic before evaporation to degeneracy. All-optical forced evaporation is performed in 5 seconds by adiabatically decreasing the power in the crossed dipole trap (CDT) in three linear ramps. This produces a Bose-Einstein condensate (BEC) comprised of order $4 \times 10^4$ rubidium-87 atoms with measured temperatures below 10~nK. We can easily produce much larger condensates, but intentionally restrict the number to give repeatability. Following evaporation, the CDT is maintained at its final evaporation power to offset the effect of gravity during subsequent experiments.

The 3D optical lattice is generated by splitting the output of a 30~W 1064~nm IPG laser into 6 independent beams. A considerable amount of power is dumped prior to splitting since only hundreds of milliwatts per beam is needed to construct the lattice. Following beam shaping and polarizing optics, a pair of counter-propagating lattice beams are projected onto the atoms from each of three orthogonal Cartesian directions. The six lattice beams all pass through their own acousto-optic modulator (AOM), which controls both the intensity and relative phase of the optical standing waves. To avoid interference between orthogonal axes of the lattice, we select opposite diffraction orders from the AOMs for the $\hat{x}$ and $\hat{y}$ axes and detune our vertical axis AOM radio frequency drive by 1~MHz relative to the other axes, so that any interference is time-averaged and can be neglected. Measured beam waists for the 3D lattice are $w_x, w_y = 95\,\mu$m and $w_z = 150\,\mu$m, where $\hat{z}$ is parallel to gravity and both $\hat{x}$ and $\hat{y}$ reside in the horizontal plane. For the 2D lattice experiments reported here, we only apply lattice beams in the $\hat{x}$--$\hat{z}$ plane. Along each lattice direction, we also pass a resonant (780 nm) probe beam for absorption imaging on a CMOS camera. This allows us to reconstruct the momentum probabilities that represent the interferometer output. 

For the experiments that follow, the atoms are subject to a net optical dipole potential $V(\bm{r}, t)=V_L(\bm{r}, t) + V_D(\bm{r})$ at point $\bm{r}=(x,z)$ and time~$t$. The lattice potential is given by
\begin{equation}
    V_L(\bm{r}, t) = \frac{V_0}{2}\left(\cos\bigl(2kx + \phi_x(t)\bigr) 
    + \cos\bigl(2kz + \phi_z(t)\bigr)\right)\,,
\end{equation}
where $k = 2\pi/\lambda$ is determined by the laser wavelength $\lambda$, and $\phi_x(t)$ and $\phi_z(t)$ are control functions of the standing wave. The depths of the lattice beams were calibrated to be the same using Kapitza-Dirac diffraction \cite{kapitza_dirac_1933, PhysRevLett.56.827}, i.e., $V_0 = 10\,E_r$, where $E_r=\hbar^2k^2/2m$ is the recoil energy, with $m$ the atomic mass. The ellipsoidal CDT potential, $V_D(\bm{r})$, has measured $x$, $y$ and $z$ harmonic trap frequencies of 37~Hz, 130~Hz, and 135~Hz respectively, as obtained via parametric heating \cite{PhysRevLett.31.1279}.  

At the end of an interferometer sequence, the trapping beams are extinguished, and absorption imaging of the atomic ensemble performed. The absorption image is sensitive to the column-integrated optical density as the probe beam passes through the cloud. Before imaging, the cloud is expanded in time of flight (TOF) for 10--14~ms so that different momentum components separate and can be discriminated. Since the lattice photons can only transfer quantized momenta, the observed diffraction orders are discrete and separated by $2\hbar k$ along each axis. Deducing the total optical density in each diffraction order provides normalized atom numbers for each momentum component. 

In the following sections we demonstrate application of this experimental method to perform 2D vector accelerometry via simultaneous Bloch oscillations and to perform a machine-designed Michelson vector atom interferometer. 

\vspace*{-1pc}
\subsubsection*{Multidimensional Bloch oscillations}
\vspace*{-0.75pc}

Although Bloch oscillations \cite{1929ZPhy...52..555B,1934RSPSA.145..523Z} were originally formulated in the context of electron transport in condensed matter systems, ultracold atoms in optical lattices have become an excellent platform for both measuring \cite{PhysRevLett.76.4508, PhysRevLett.100.080404} and inducing Bloch oscillations, leading to sensitive inteferometry~\cite{Gebbe2021, PhysRevLett.102.240403, PhysRevA.85.013639}. The eigenstates of the infinite periodic lattice potential are characterized by the non-negative integer Bloch band index,~$n$, and continuous bounded quasimomentum,~$q$. The quasimomentum is restricted to the first Brillouin zone of the periodic potential $q\in[-\pi\hbar/d, \pi\hbar/d]$, where $d=\lambda/2$ is the lattice spacing. Following BEC production, the 2D lattice beams were adiabatically ramped on in 1 ms, at which point the atoms resided in the ground Bloch state with zero quasimomentum, i.e.,  $\ket{n=0, q=0}$. When an impulse force~$F$ is subsequently applied, $q$ evolves at time $t$ according to $q(t) = q(0) +  Ft$. For a constant positive force that is not strong enough to result in interband transitions, the quasimomentum increases until it reaches the edge of the first Brilloiun zone, $q=\hbar k$, at which point the atoms resonantly lose $2\hbar k $ of momentum due to Bragg reflection and reset their quasimomentum to $q=-\hbar k$. From this point they are accelerated again, leading to a repeating cycle with a Bloch period $\tau_b = 2\hbar k/F$.

\begin{figure}[!ht]
    \centering
    \includegraphics[width=0.48\textwidth]{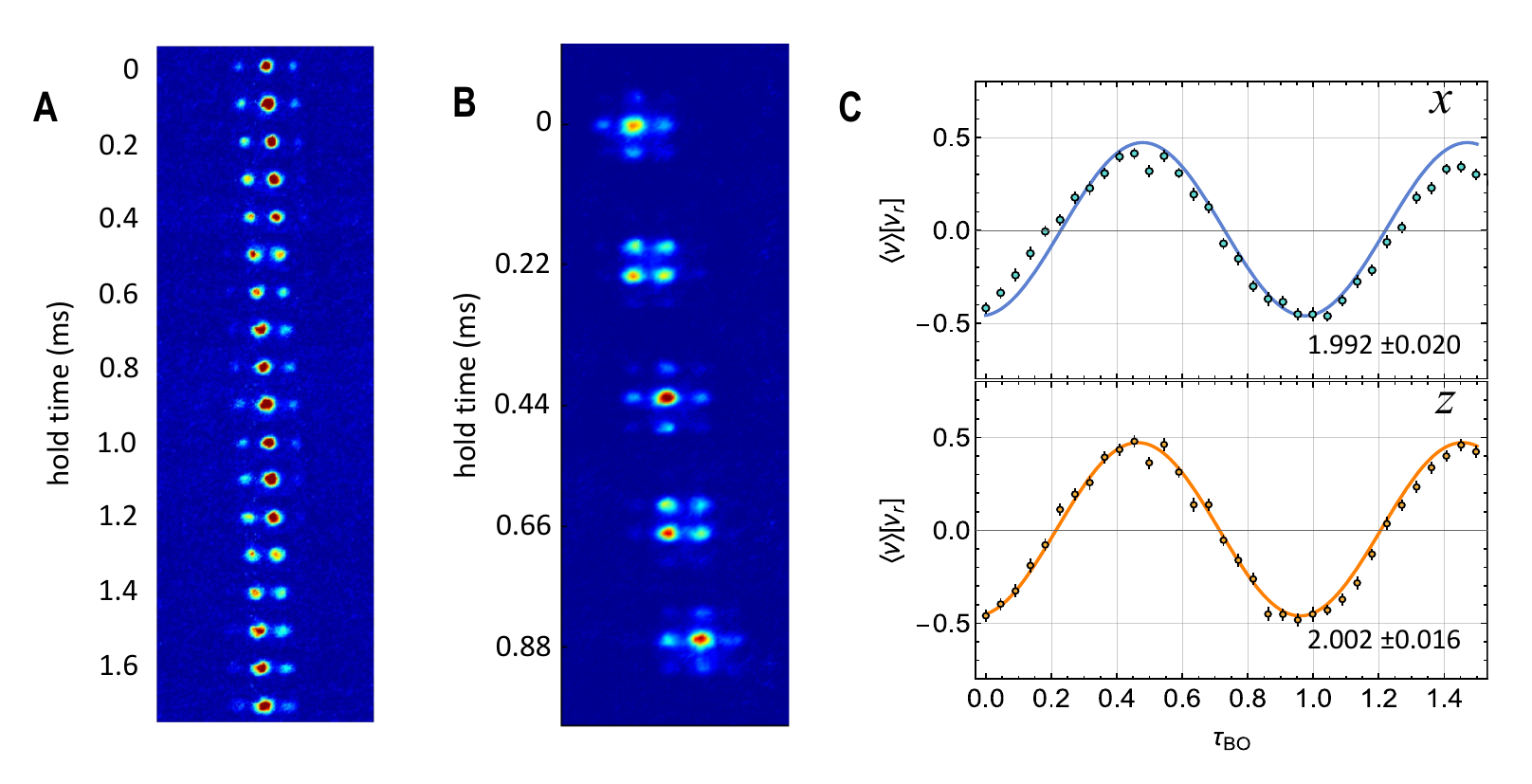}
    \caption{\textbf{Bloch Oscillations:} For different hold times, raw image slices of the observed momentum diffraction orders stacked vertically. In each image, the separation between momentum peaks is~$2\hbar k$. (A) Oscillations in a 1D optical lattice with applied acceleration 1$g$. (B) Simulataneous Bloch oscillations in a 2D optical lattice with applied accelerations of $2g$ in both dimensions, image slices taken at every half Bloch period. (C) Associated fits to the 2D Bloch oscillations for the $x$-axis (upper) and $z$-axis (lower) and the extracted acceleration values given in $g$. The errorbars denote the standard deviations of the population fractions for five independent experimental runs.   
    }
    \label{fig:BO Fringes}
\end{figure}

In Bloch oscillation experiments, atoms are often confined in a vertical optical lattice in which gravity is used to exert a constant (but fixed) force, see Fig.~\ref{fig:BO Fringes}A. By measuring the mean velocity of the atomic ensemble over many Bloch oscillations, it is possible to extract the magnitude of the local gravitational acceleration. Several lattice experiments have successfully observed hundreds to thousands of Bloch periods to obtain sensitive measurements of gravity \cite{PhysRevLett.106.038501,  PhysRevLett.92.230402}. Since our lattice axes are derived from independent beams, it is possible to apply a large range of accelerations to the confined atoms. For each lattice axis, this involves executing a linear sweep of the frequency difference of the pair of interfering lasers. Applying concurrent linear frequency ramps to two lattice axes simulates a vector force, and this allows us to observe the effects of both the direction and magnitude of the applied acceleration in the observed Bloch oscillation patterns~\cite{PhysRevA.67.063601,Witthaut_2004}.

In Fig.~\ref{fig:BO Fringes}B, we present characteristic absorption images of the 2D momentum state following TOF for varying hold times in the optical lattice. An acceleration of 2$g$ was applied to each axis of the lattice simultaneously, and subsequent 2D Bloch oscillations were observed. Fig.~\ref{fig:BO Fringes}C shows the normalized momentum state populations with sinusoidal fits. The extracted values of  $a_x=(1.992\pm0.020)g$ for the $x$-axis and $a_z=(2.002\pm0.016)g$ for the $z$-axis are in good agreement with those applied. Although for brevity we give only one example here, measurement of both Bloch periods permits determination of the absolute magnitude of the force $|\Vec{F}| = m\sqrt{a_x^2 + a_z^2}$, and the relative orientation in the $\hat{x}$--$\hat{z}$ plane via: $\theta = \text{arctan}(a_z/a_x)$.

\vspace*{-1pc}
\subsubsection*{Vector atom  accelerometer}
\vspace*{-.75pc}

\begin{figure}[ht]
    \includegraphics[width=0.42\textwidth]{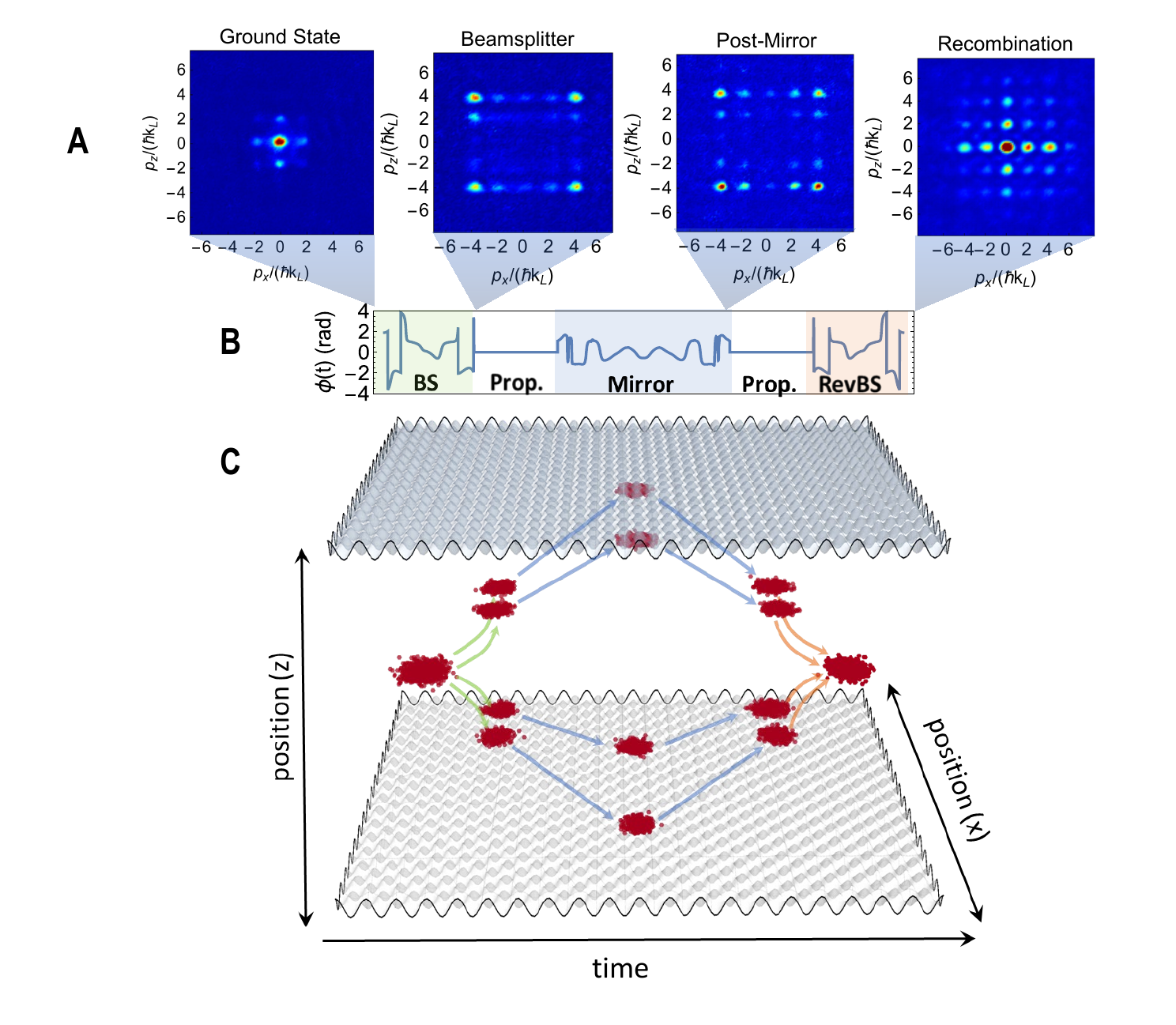}
    \caption{\textbf{Vector atom interferometer:} Experimental images of a Michelson interferometer sequence in 2D. (A) Observed momentum components at the start of the sequence, after the beam splitter, after the mirror and at the end of the sequence after recombination. Ideally the first and last images should be the same, as should the images before and after the mirror (since positive and negative momenta simply exchange.) (B) The control functions that were applied to both $x$ and $z$ dimensions. Here a propagation stage is included for visualization, but omitted in the two case studies we show later. (C) Illustration of the BBI system, which involves an omnipresent optical lattice, with the control functions applied during the beam splitter and mirror components, and propagation occurring in the lattice conduction band. The sequence shows the BEC splitting into four 2D momenta (green arrows), reflection after propagation (blue arrows), and the reverse beam splitter providing recombination (orange arrows) to form the 2D interference pattern from which the inertial phase is revealed.
    }
    \label{fig:machzehnder}
\end{figure}

We experimentally demonstrate a 2D Michelson interferometer by stitching learned beamsplitter, mirror, and recombination components into a single control function for each axis. High fidelity control protocols for each component were produced by both quantum optimal control (QOC) \cite{PRXQuantum.2.040303,PhysRevA.105.032605,10015539} and reinforcement learning (RL) \cite{RL_KapitzaOscillator, RL_manyBody, PhysRevResearch.3.033279}. Since the lattice dynamics is approximately separable between dimensions, a 2D interferometer can be realized by the simultaneous application of the protocols designed for 1D to each axis, see Fig.\ref{fig:machzehnder}. The control functions used here are the same as the QOC protocols reported in Ref.~\cite{10015539}. 

The start of the interferometer sequence is the beamsplitter, targeting the Bloch state transformation $$\ket{n=0, q=0}\longrightarrow\ket{n=3, q=0}$$ and thereby exciting atoms from the ground state of the lattice where they are deeply confined into the conduction band where they can propagate. This is due to the fact that $\ket{n=3, q=0}$ for a $10E_r$ deep lattice is to good approximation an equal superposition of $\pm4\hbar k$ momenta---enough kinetic energy (i.e., 16$E_r >$ 10$E_r$) to efficiently transport through the lattice potential. In 2D, the beamsplitter creates a superposition of four wave packet components that spatially separate in a square pattern.

We show that we can achieve a sensitive interferometer without free propagation stages in between the beamsplitter and mirror components. This enables us to quantify an extremely simple system, with the alternate paths of the interferometer spatially separating by only a few wavelengths of the lattice light. This choice is motivated by the fact that this is the first ever demonstration of BBI of its kind. For this case, the entire interferometer sequence is completed in a total of 460~$\mu$s, with the control functions for the beamsplitter and mirror requiring 118~$\mu$s and 236~$\mu$s respectively. It will be straightforward in future work to scale to longer times by including propagation stages. 

For a lattice depth of $10E_r$, it is necessary to monitor the quantized momentum components in each dimension over roughly $7$ values $[-6\hbar k, -4\hbar k, \ldots, 6\hbar k]$. This means that there is an essential difference between our BBI system and conventional atomic interferometers that discriminate just two states. Here the readout from the 2D Michelson interferometer is multichannel with $7\times7=49$ relative weights in a 2D momentum grid. External forces applied during the interferometer sequence alter the momentum probabilities due to the differential phases that are accumulated. The resulting pattern represents a fingerprint that potentially allows one to recognize the applied inertial force. Inertial forces are introduced by linear frequency ramps of the lasers along each axis, as was demonstrated for Bloch oscillations. 

\vspace*{-1pc}
\subsubsection*{Case study 1: Varying amplitude, constant direction}
\vspace*{-.75pc}

\begin{figure}[!htb]
    \centering  \includegraphics[width=0.48\textwidth]{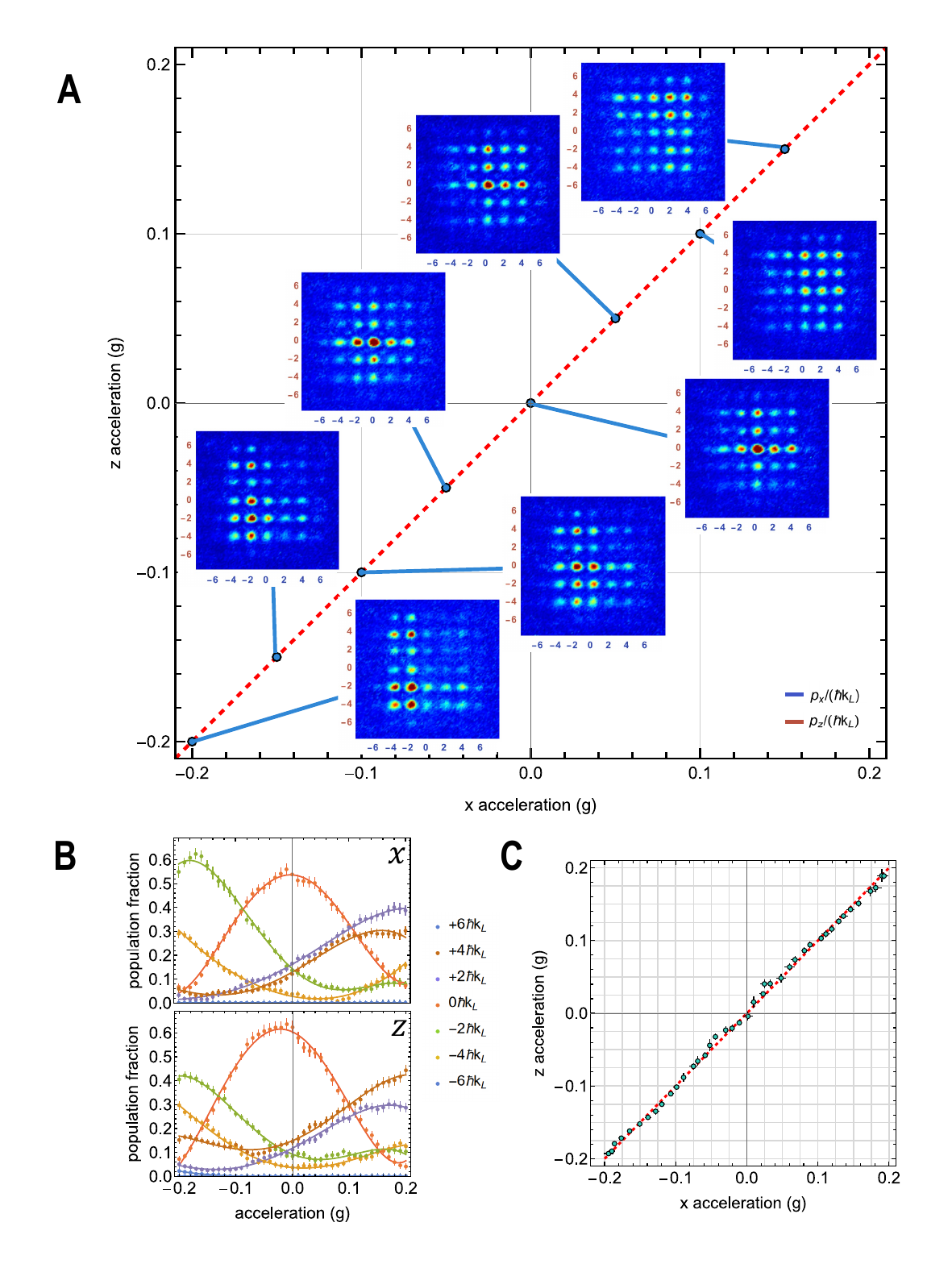}
    \caption{\textbf{2D Michelson amplitude scan:} (A)  Momentum diffraction images for a range of acceleration vectors with $a_x=a_z$ but varying in amplitude, both positive and negative. (B) Empirical model made by projecting the momentum diffraction scans onto the component directions $x$ and $z$, giving a mean value and errorbar from the repeated measurements, and fitting a B-spline. (C) Multiparameter estimation from the minimum least squares fit to the empirical model. Errorbars arise from the uncertainty due to the variation in the reconstruction from multiple experimental runs.}
    \label{fig:Linear Scan}
\end{figure}

We apply equal accelerations to the $x$ and $z$-axes of the lattice but vary the amplitude, both positive and negative. A total of 41 pairs of accelerations were investigated covering the range from $-0.2g$ to $0.2g$. For each pair of accelerations, 10 experimental images were taken and averaged to extract multidimensional momentum populations for each component of the momentum grid. In Fig.~\ref{fig:Linear Scan}A, momentum diffraction images for $a_x = a_z$ are displayed. We draw attention to how the momentum components with the greatest populations seen in red in Fig.~\ref{fig:Linear Scan}A evolve as the acceleration values are scanned. For $a_x=a_z=-0.2g$, atoms overwhelmingly occupy the lower left quadrant of the momentum state pattern. As the magnitude of the applied acceleration is decreased $(a_x,a_z\rightarrow0$), atoms primarily occupy momentum clouds around the center ($0\hbar k$, $0\hbar k$). As the acceleration approaches its maximal positive value, the atom populations are observed to shift to the upper right quadrant.
 
Figs.~\ref{fig:Linear Scan}B and \ref{fig:Linear Scan}C demonstrate the quantitative performance for parameter reconstruction. In Fig.~\ref{fig:Linear Scan}B, we have assumed the Hamiltonian for the device to be approximately separable in  dimensions, and integrated our images over all but one of the dimensions in turn for each acceleration, giving rise to marginal probability distributions for each axis. The discrete acceleration values that were measured can then be interpolated to give a continuous reference for each component of a 7-dimensional momentum probability vector. This method provides a simple data-driven empirical model that we can use to calibrate the device for sensing an unknown acceleration. Any 2D image can be marginalized into two 1D distributions and referenced against the empirical model to infer the vector acceleration. For each image, we minimize the Euclidean distance of its marginals to the empirical model, giving the estimated acceleration components. In Fig.~\ref{fig:Linear Scan}C, we show this procedure works well. Since each acceleration point was repeated 10 times, we also obtain errorbars. 

\vspace*{-1pc}
\subsubsection*{Case study 2: Varying amplitude, constant magnitude}
\vspace*{-.75pc}

\begin{figure}[!htb]
    \centering
    \includegraphics[width=0.48\textwidth]{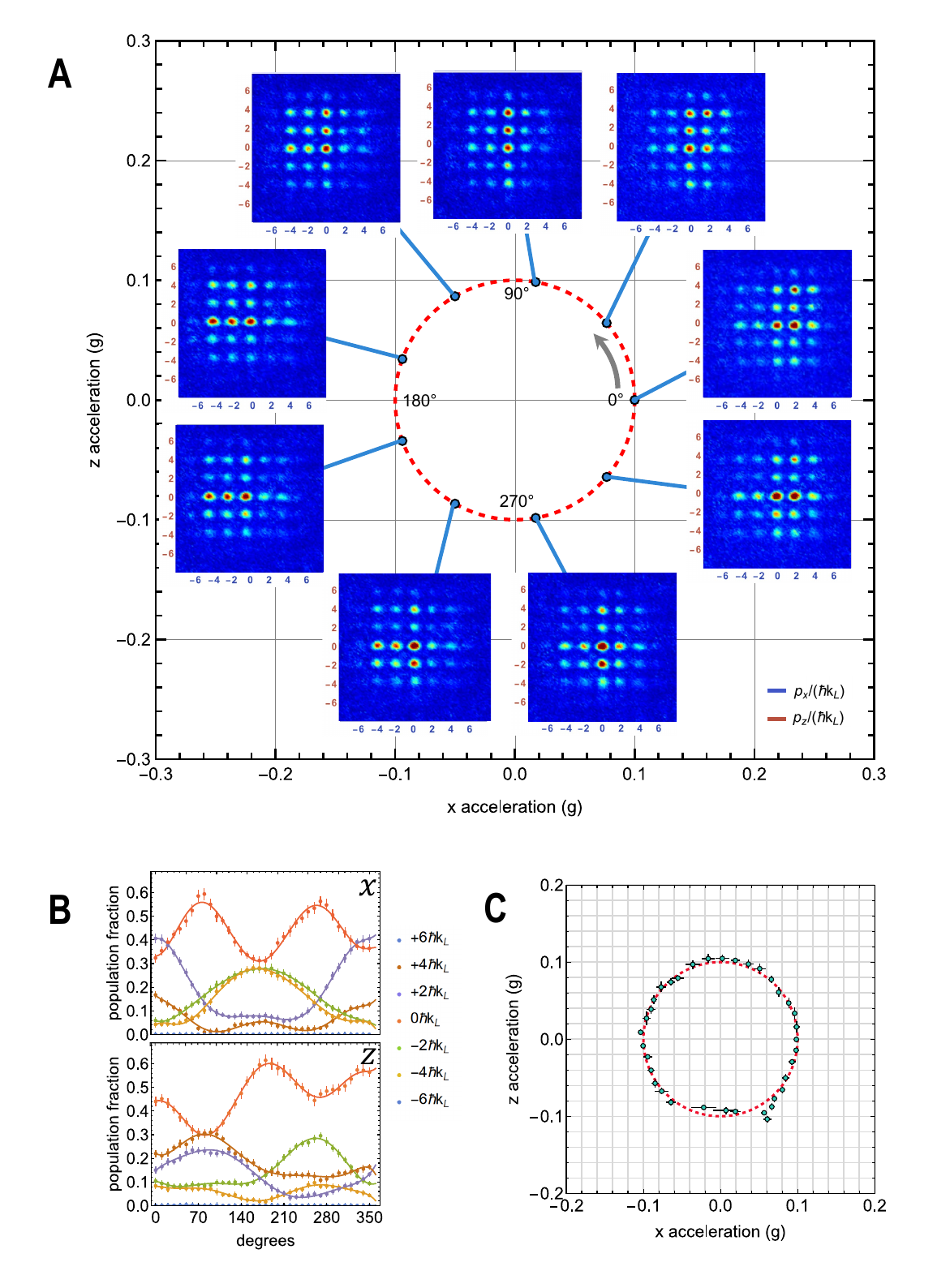}
    \caption{\textbf{2D Michelson polar scan:} (A) Momentum diffraction images for a range of acceleration vectors with $|a_x|=|a_z|=0.1g$ but varying the polar angle and thus the acceleration direction. (B)~The empirical model that results from projecting each image onto both dimensions and fitting B-splines to the momentum fringes. (C) Multiparameter estimation from the minimum least squares fit to the empirical model recorded in Study 1, i.e., given in Fig.~\ref{fig:Linear Scan}B,  with multiple experiments taken at each point to construct errorbars.}
    \label{fig:Polar Scan}
\end{figure}

We show that the 2D Michelson interferometer can discriminate the direction of applied forces. We fix the magnitude of the applied accelerations and vary the polar angle of the acceleration vector in the $x$-$z$ plane. Applied accelerations were of the form,
\begin{eqnarray}
        a_x &=& |a_x|\cos\theta \nonumber\\
        a_z &=& |a_z|\sin\theta
\end{eqnarray}
where the magnitude of the acceleration was kept fixed $|a_x|=|a_z|=0.1g$, and the polar angle was scanned for $\theta \in [0, 2\pi]$ in increments of $\pi/20$. For an acceleration of this magnitude, the central clouds of the momentum grid remain populated, however, off-center momentum components become pronounced as the polar angle is varied. While many of the momentum state components do not completely vanish, large population contrast in the momentum grid images is observed over a $\pi$ phase change in the direction of the applied force, as shown in Figs.~\ref{fig:Polar Scan}A. As for the previous study, we can construct the interpolated marginal distributions for each axis, shown in Fig.~\ref{fig:Polar Scan}B. In Fig.~\ref{fig:Polar Scan}C we have carried out the analogous procedure to Fig.~\ref{fig:Linear Scan}C, including the best fit and uncertainties, except that the Euclidean distance calculation uses the empirical model Fig.~\ref{fig:Linear Scan}B, which was constructed using the case study 1 data. Since this reference is created from a distinct data set, the reconstructed vector force estimate is unbiased. When combined with the previous case study, these results indicate that we are able to perform high-quality sensing of an arbitrary unknown vector acceleration.

\vspace*{-1pc}
\subsubsection*{Analysis: Bayesian Reconstruction \& Averaging}
\vspace*{-.75pc}

Having established a method for extraction of multiple parameters from single absorption images, the next question we consider is how to optimally combine information from multiple experimental runs to determine quantitative limits for the overall short-term sensitivity of the device. In traditional atom sensors, i.e., light-pulse interferometers, this is usually a straightforward process since the sensitivity can be obtained from a model fit of sinusoidal fringes to the measured data from multiple experimental runs. However, for BBI interferometers, the multidimensional diffraction pattern contains many output momentum ports thereby prohibiting a simple oscillating solution. Instead, the estimation task can be articulated in terms of the language of Bayesian inferencing for the measurement of parameters. This implies a need to develop a forward model for Bayesian updating that goes beyond the simple fringes used in a typical atom interferometer. The completely unbiased approach would be to use an ab~initio theory of the experiment for the forward model, but this would make the calibration explicitly dependent on the degree of uncertainty of inputs, i.e.,~details of the apparatus and included physics. 

As in the two studies, we instead take a data-driven approach to the analysis and use the parameterized fit given in Fig.~\ref{fig:Linear Scan}B as our reference model. This allows combining experimental shots by Bayesian inferencing, which updates the probability for a hypothesis of parameter values as more data is accumulated. Consider first a single atom passing through the BBI sequence, and a projective measurement made at the output of its momentum. The measurement records a specific value for a bin index $m$, in our case $m$ can take on 1 of 49 possible values corresponding to the enumerated diffraction peaks in the momentum grid image. The resulting posterior probability distribution is found by Bayesian updating,
\begin{equation}
    P(a_x, a_z | m) = {\cal N}P(m | a_x, a_z) P(a_x, a_z)
    \label{eq:bayes}
\end{equation}
where ${\cal N}$ is a normalization coefficient, $P(a_x, a_z)$ is called the prior, and $P(m|a_x, a_z)$ is the forward model. Since we have about $4\times10^4$ BEC atoms, if we were able to detect every atom, and if each atom measurement was truly independent, we could in principle iterate this equation tens of thousands of times. This independent atom limit for the posterior probability is known as the standard quantum limit in metrology. 

However, as is typical in quantum gas experiments of this type, we are not able to detect every atom due to technical noise in the imaging system. Even so, we can quantify the number of independent trials $N_{\rm trial}$ that can be extracted from each image by performing many experimental runs and comparing the mean value in each bin to the fluctuations shot to shot. In probability theory, for $N_{\rm trial}$ independent trials, each of which can produce a result in one of $m$ categories, the distribution is a multinomial, a generalization of the binomial distribution. When we fit the multinomial to our data set, we find $N_{\rm trial}$ to be much smaller than the number of atoms. This is due to probe intensity fluctuations, as well as the quantum efficiency inherent in photodetection and in photon absorption as the light propagates through the ensemble. 

\begin{figure}[!htb]
    \centering
    \includegraphics[width=0.48\textwidth]{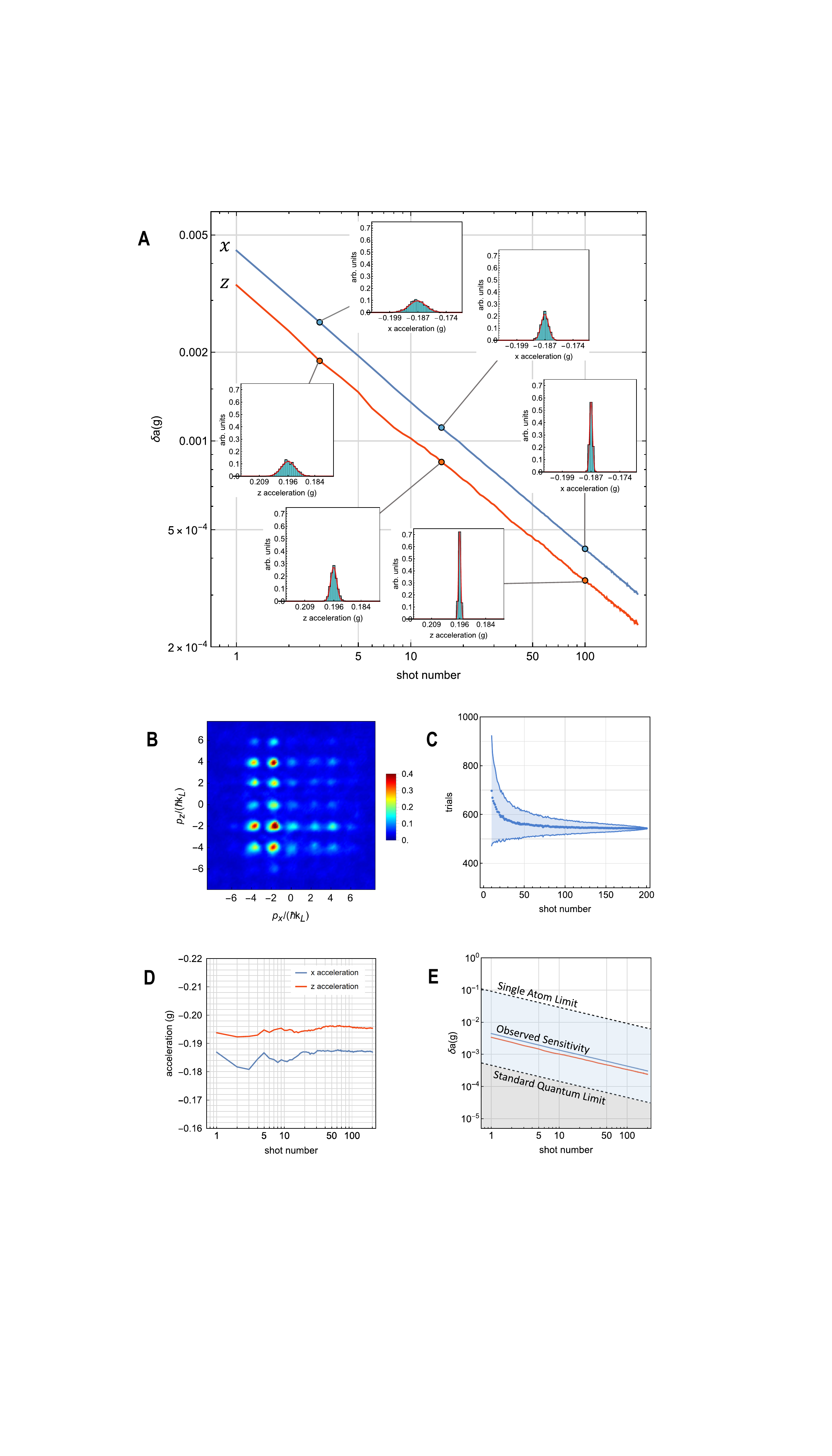}\caption{\textbf{Sensitivity:} Parameter estimation and associated uncertainty from 200 repeated experiments with the same $-0.2g$ acceleration applied in both dimensions. (A) Standard deviation $\delta a$ of the posterior probability distribution after sampling $N_{\rm trial}$ times from the observed normalized experimental momentum probability. Histograms are of the parameter estimator (using the mean-value of the posterior) from many random samples of size 3, 15, and 100 taken from the 200 shot experimental image pool. (B) Average experimental momentum probability image. (C) Predictor of the number of independent trials that can be extracted from a single image found from the shot-to-shot variation between bins of different images and fitting to a multinomial distribution. Convergence with shot number is apparent to an asymptotic value of approximately $N_{\rm trial}=532$. (D) Reconstructed value of the vector acceleration from the mean value of the posterior distribution. (E) Comparison of the observed sensitivity in (A) with the prediction bounds from the classical Fisher information of the reference model at $-0.2g$ for both a single atom and for the standard quantum limit with $4\times10^4$ atoms. }
    \label{fig:Bayesian}
\end{figure}

In Fig.~\ref{fig:Bayesian}A, we show the sensitivity as data from 200 experimental runs are accumulated, a process that takes a few hours given our current cycle time. For this example, we applied a fixed acceleration of $-0.2g$ in both dimensions and repeated shots, i.e., performed identical experiments. We show the resulting standard deviation $\delta a$ of the posterior probability distribution for both axes. Taking many different samples of size 3, 15, and 100, from the 200-image pool for each axis, we histogram the mean value of the resulting posterior distribution. This gives a complementary measure of the device sensitivity with the width of the histograms for distinct data sets in good agreement with $\delta a$. We show the extraction of the number of independent trials $N_{\rm trial}$ in Fig.~\ref{fig:Bayesian}C; as we take different numbers of shots, $N_{\rm trial}$ converges rapidly to a value of approximately 532. This is much less than the actual atom number, which we know independently from total absorption and scaling law dynamics in TOF~\cite{PhysRevLett.77.5315}. In Fig.~\ref{fig:Bayesian}D, we show the mean value of the posterior probability, i.e., the multiparameter estimation, on a semilog scale. We calculate the classical Fisher information from the reference model for the applied acceleration, and show in Fig.~\ref{fig:Bayesian}E the bounds these give for the sensitivity for a single atom measurement and for the standard quantum limit that corresponds to $4\times10^4$ atoms. We find there is scope to improve the device sensitivity by improvements in the quantum efficiency of atom detection, which should allow us to measure closer to the standard quantum limit. 

We emphasize that the resulting sensitivities that we measure of $\approx10^{-4}g$ are impressive given that the interferometer spatial scale is only a few sites of the the lattice, and the operation time is less than half a millisecond. Since the short-term sensitivity scales with the enclosed area in space-time, $\delta a$ decreases in general as the square of the propagation time between components. As this interferometer is scaled up to times that are realistic to achieve in near future experiments (10--100~ms), with corresponding millimeter spatial scales of the interferometer footprint, the sensitivities achieved will enter into a technologically relevant domain. For example, with a propagation time of 100~ms, and the same atom number and integration time for averaging down as was performed here,  the predicted sensitivity would be $1.3\times10^{-9}g$.

\begin{figure}[!htb]
    \centering
    \includegraphics[width=0.48\textwidth]{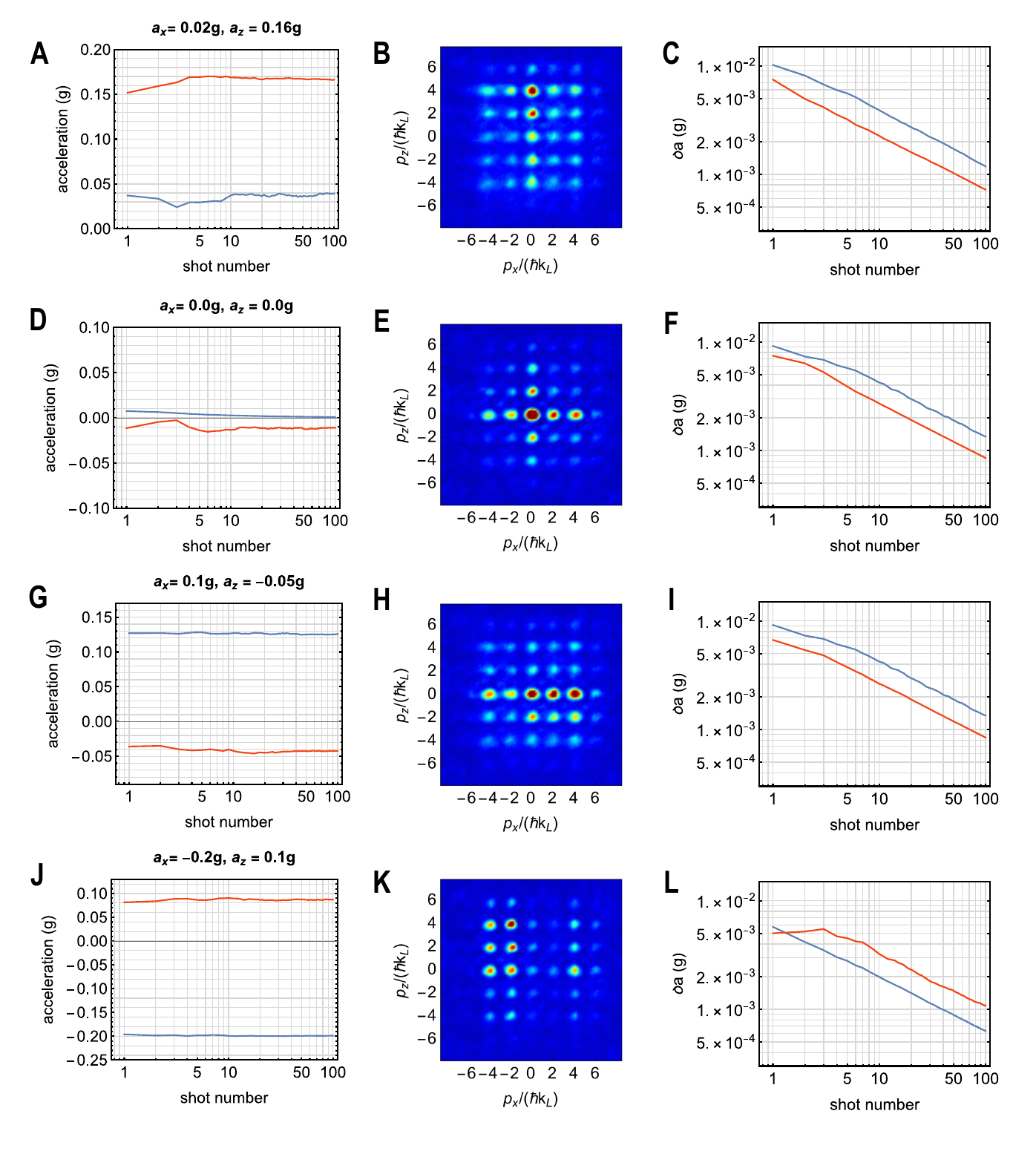}
    \caption{\textbf{Vector acceleration measurements:} Bayesian reconstruction for 2D acceleration parameter estimation, each row a different example. The columns left to right give the mean value of the posterior probability distribution as the parameter estimator, the observed momentum diffraction images giving the fingerprint from which the parameters are derived, and the standard deviation of the posterior probability giving the sensitivity. For each example, 100 identical experiments (shots) were performed.}
    \label{fig:100runvar}
\end{figure}

In Fig.~\ref{fig:100runvar}, we run experiments for 4 different examples of the vector acceleration, showing the mean value estimator (first column), the momentum grid images that were observed (second column), and the sensitivities (last column). In each case, 100 experimental shots were taken. This provides a good overview of the how the device works in practice; it is noticeable how different the image patterns are for different values of the vector acceleration, and that across a broad spectrum of acceleration parameters, high measurement sensitivity can be achieved.

\vspace*{-1pc}
\subsubsection*{Discussion \& Conclusion}
\vspace*{-.75pc}

In this paper, we have presented two novel multidimensional vector atomic accelerometers, capable of sensing both the magnitude and direction of applied inertial signals. A natural extension to a 3D optical lattice would permit the device to be configured as a general-purpose accelerometer, gyroscope, or gravity gradiometer, on demand. This is because all of the control functions for BBI are software based and can be combined in programmed sequences that do not require hardware changes. With the ability to program, our device could realize appealing scenarios that sense accelerations along three axes, rotations in arbitrary planes, or even the 5 independent components of the gravity gradient tensor in a single instrument. This flexibility, coupled with the rapid convergence of the posterior probability to extract applied accelerations with aliasing protected Bayesian reconstruction, positions BBIs as an attractive field-deployable technology that may overcome many challenges faced by traditional atom interferometers.

While we stress our optical lattice interferometer results have thus far been for small interferometers, we believe interferometer sensitivities can be increased by many orders of magnitude. This will involve improvements to the spatial quality of our lattice beams and deployment of an active phase control servo loop which could enhance sensitivity via extended interferometer propagation times. Ongoing upgrades to our evaporation sequence and absorption imaging are currently being implemented to increase both the data rate and the information contained within each shot, allowing us to measure closer to the standard quantum limit. These experimental modifications will be paired with the exploration of novel accelerometer geometries as we advance this platform to scientifically relevant sensitivities. Additionally, since our interferometer does not rely upon coupling of internal states, one intriguing prospect is that our Bloch-band interferometer offers improved immunity to stray electric and magnetic fields commonplace to dynamic environments. This technology opens an entirely new class of atomic interferometers, where the atoms can be confined throughout the entirety of their interrogation, kept within a single internal state, and the functionality programmed on demand, to meet evolving sensing objectives \cite{saywell-2023}.  

\vspace*{-1pc}
\subsubsection*{Acknowledgements}
\vspace*{-.75pc}

We would like to thank John Wilson, Shah Saad Alam, Malcolm Boshier, Ceren Uzun, James Thompson, Jun Ye, John Kitching, and the NASA Quantum Pathways Institute for many helpful discussions. This research was supported in part by NASA under grant number 80NSSC23K1343, by NSF OMA 1936303, NSF PHY 1734006, NSF OMA 2016244, and NSF PHY 2207963.

\bibliography{ref}


\renewcommand{\theequation}{S.\arabic{equation}}
\setcounter{equation}{0}
\renewcommand{\thefigure}{S\arabic{figure}}
\setcounter{figure}{0}

\end{document}


\section*{Supplementary Material}

\vspace*{-1pc}
\subsubsection*{BEC Production}
\vspace*{-.75pc}
We produce a Bos-Einstein Condensate, or BEC in a similar manner to that described in [cite arxiv paper]. However, in-between the results presented in [cite arxiv paper] and the results described here, the experimental apparatus has been completely rebuilt in order to incorporate a three-dimensional lattice.  The experimental system is still constructed around a ColdQuanta physics station which houses a custom double magneto-optical trap (MOT) cell, and the details behind obtaining a MOT along with performing sub-Doppler cooling remain the same. After polarization-gradient cooling (PGC) we have around $6\times 10^8$ atoms at about 20~$\mu$K. Atoms are then loaded into a crossed-dipole trap (CDT), which is constructed from a 1064nm 30W IPG laser. The second dipole beam is generated by re-passing the first beam through the science cell, with the beams intersecting at $32^{\circ}$and each having 55~$\mu$m  waists. Evaporation in the CDT follows that presented in [cite arxiv paper]. Atoms are loaded into the CDT in the $F=1$ state and after a 50ms hold, evaporation is performed adiabatically by linearly ramping down the CDT power in three successive stages. Our total optimized evaporation sequence takes approximately 5s, after which we typically produce BECs with as many as $10^5$ atoms with temperatures below 20 nK.

\vspace*{-1pc}
\subsubsection*{3D Lattice}
\vspace*{-.75pc}
As seen in Figure 1A of the main manuscript, our 3D lattice is constructed of 3 orthogonal lattice beams. An internal mirror angled at $4^{\circ}$ allows for implementation of the vertical lattice beam even with optical access restricted to the top of the cell. In the 3D lattice scheme, each lattice beam passes through its own respective acousto-optic modulator (AOM) which allows one to implement a relative phase to the lattice by simply introducing a frequency difference between each pair of counter-propagaing lattice beams via the AOMs. Shaking protocols are then applied in the same method described in [cite arxiv paper]. Essentially for each pair of lattice beams we apply a phase modulation by updating the RF frequency driving one of the lattice AOMs. To do this we utilize an Agilent 33622A arbitrary waveform generator and typically sample the modulation waveforms of a shaking protocol with 50 ns resolution. When there is no acceleration applied, the AOM driving the second lattice beam is driven from the same waveform generator, but with the constant carrier frequency. To apply an acceleration signal to the atoms, the second AOM is instead driven with a linear frequency sweep away from its carrier. The frequency sweep and the start of the shaking sequence for the atom interferometer are both simultaneously triggered following initial lattice load. Lattice load and shaking protocols are applied simultaneously for multiple dimensions of the lattice. The lattice load is adiabatic and preformed over 1ms, and the x-axis lattice beam used to produce the data presented in the main manuscript is co-linear with the first pass of the CDT.

\vspace*{-1pc}

\newpage

\clearpage